\documentclass[11pt,a4paper]{article}

\usepackage[utf8]{inputenc}
\usepackage[T1]{fontenc}
\usepackage{graphicx}
\usepackage[dvipsnames]{xcolor}
\usepackage{caption} 
\usepackage{multirow}
\usepackage{amsmath}

\usepackage{amssymb}
\DeclareMathOperator*{\argmin}{arg\,min}
\usepackage{amsthm}

\usepackage{graphicx}
\usepackage[colorinlistoftodos]{todonotes}
\usepackage{csquotes}
\usepackage[colorlinks=true, linkcolor=black, urlcolor=black, citecolor=black]{hyperref}
\usepackage[a4paper]{geometry}
\usepackage[]{dcolumn}
\usepackage{rotating}
\usepackage[]{algorithm2e}

\usepackage{nameref}
\usepackage{indentfirst}
\usepackage{comment}
\usepackage{rotating}

\usepackage{textcomp}
\usepackage{setspace}
\usepackage{geometry}
\usepackage{float}
\usepackage[normalem]{ulem}
\usepackage{cancel}

\usepackage{booktabs}
\usepackage{siunitx}
\usepackage{authblk}
\title{Synthetic Difference in Differences for Repeated Cross-Sectional Data}

\author[1,*]{Yoann Morin}

\affil[1]{CESAER UMR1041, INRAE, Institut Agro, Université Bourgogne Franche-Comté}
\affil[*]{Corresponding author: yoann.morin@protonmail.com}
\date{}                     
\usepackage{caption}
\usepackage{lmodern}

\newtheorem{assumption}{Assumption}

\usepackage{lscape}

\newcommand{\homega}{\hat{\omega}}
\newcommand{\hlambda}{\hat{\lambda}}
\newcommand{\pre}{{\rm pre}}
\newcommand{\post}{{\rm post}}
\newcommand{\sdid}{{\rm sdid}}
\newcommand{\mmr}{\mathbb{R}}
\newcommand{\ccc}{{\rm co}}
\newcommand{\ttt}{{\rm tr}}
\newcommand{\cb}[1]{\left\{#1\right\}}

\newcommand{\sumit}{\sum_{i=1}^N\sum_{t=1}^T }

\usepackage{natbib}
\setcitestyle{authoryear,open={(},close={)}}

\usepackage[toc,title,page]{appendix}

\begin{document}
\maketitle

\begin{abstract}
The synthetic difference-in-differences method provides an efficient method to estimate a causal effect with a latent factor model. However, it relies on the use of panel data. This paper  presents an adaptation of the synthetic difference-in-differences method for repeated cross-sectional data. The treatment is considered to be at the group level so that it is possible to aggregate data by group to compute the two types of synthetic difference-in-differences weights on these aggregated data. Then, I develop and compute a third type of weight that accounts for the different number of observations in each cross-section. Simulation results show that the performance of the synthetic difference-in-differences estimator is improved when using the third type of weights on repeated cross-sectional data.
\end{abstract}

\noindent \textbf{JEL Classification}: C15, C29

\bigskip

\noindent \textbf{Keywords}: causal inference, difference-in-differences, synthetic difference-in-differences, latent factor models, repeated cross-sections

\newpage

\section{Introduction}
The synthetic difference-in-differences (SDiD) method (\citealp{arkhangelsky_synthetic_2021})  combines features from difference-in-differences and synthetic control methods. As in synthetic control methods, it computes weights that match pretreatment outcomes between the treated and the control groups. This reduces the need for the parallel trend assumption, often hard to verify in practice (\citealp{roth_pretest_2022}). As in difference-in-differences methods, it controls for unit-specific shifts and can include covariates.\\

This method is based on panel data, yet, repeated cross-sectional data, where each unit is observed only once, is often the only available data. For example, the evaluation of rent control policies often relies on online listing portals data that are repeated cross-sections where cities are different groups and dwellings are the individuals.\\

In this paper, I adapt the SDiD method to repeated cross-sectional data. I focus on the case where each unit belongs to the same group across time and where the treatment is applied to one or more group. Once data are aggregated, unit and time weights are computed for each group using the SDiD method. Then, a third type of weight is computed to account for the different number of observations in each group-period. Using a simulation study, I show that this estimator (RC-SDiD) performs better in terms of bias, standard deviation and RMSE that the SDiD one when the number of observations differs in each group-period.\\

In section \ref{sec:rcsdid}, I present the synthetic difference-in-differences estimator for repeated cross-sectional data. Then, I demonstrate the properties of the RC-SDiD estimator in a simulation study in section \ref{sec:simul}. Section \ref{sec:conclusion} concludes.

\section{The RC-SDiD estimator}\label{sec:rcsdid}

Consider a dataset with $T$ independent cross-sections, $t=1,...,T$. There are $K$ groups: the first $K^\ccc$ are the control groups and the last $K^\ttt = K - K^\ccc$ the treated groups. Each group is observed for every period in the sample, but all individuals are not observed for each $t$. Thus, the individual index depends on the cross-section, such as $i(k,t)$ is an individual observed at time $t$ in the group $k$, and its outcome is $Y_{i(k,t), t}$. $N_{k,t}$ is the number of observations in group $k$ at time $t$ and $N$ the total number of individuals. The treatment is assigned to groups $K^\ttt$ at time $t \geq T_\post$.\\

We make two assumptions that differ from \citet{arkhangelsky_synthetic_2021}.
\begin{assumption}[Sampling]
\label{as:samp}
Conditional on $T=t$, the repeated cross-sectional data $\{Y_{i(k,t), t}\}_{i=1}^{N}$ is an independent and identically distributed sample.
\end{assumption}
Assumption \ref{as:samp} imposes that we observe iid draws from the underlying population. However, we do not restrict the total number of observations in each cross-section to be random. In the Monte Carlo study, we test for different levels of correlation between unobservable factors and the number of observations in a cross-section.\\

\begin{assumption}[Untreated potential outcome model]
\label{as:ifeg}
The model that generates potential outcomes for the untreated can be written:
$$ Y_{i(k,t), t} (0) =  \alpha_k + \beta_t + \Lambda_k^{'} f_t + \varepsilon_{i(k,t), t} $$
\end{assumption}
\noindent where $Y_{i(k,t), t} (0)$ is the outcome of untreated observations. $\Lambda_k$ is a vector of group factor loadings and $f_t$ a vector of time factors. Both are of size $r$ such as: $\Lambda_k = (\Lambda_1, \Lambda_2, ...,\Lambda_r)^{'}$ and $f_t = (f_1, f_2, ..., f_r)^{'}$. $\alpha_k$ is a group level fixed effect and $\beta_t$ is a time fixed effect.\\
Assumption \ref{as:ifeg} means that the factors are at the group level, as the weights we compute.\\

I apply the SDiD method on aggregated data. $Y_{i(k,t), t}$ is aggregated as:
\begin{equation}
\label{eq:aggregate}
\bar{Y}_{k,t} =  \frac{1}{N_{k,t}} \sum_i Y_{i(k,t), t}  
\end{equation}
where $\bar{Y}_{k,t}$ is the average of the outcome in group $k$ at time $t$.\\
Those aggregated data are used to compute the weights $\homega^{sdid}$ and $\hlambda^\sdid$ presented in \citet{arkhangelsky_synthetic_2021} and the weights $\nu_{k,t}^{RC}$ specific for repeated cross-sectional data we introduce in this paper. The first weights to estimate are similar to those used in synthetic control methods (\citealp{abadie_synthetic_2010}). They match the outcome of the treated group with a combination of the outcome in the control groups in the pre-treatment period. They also include a group-specific shift that is accounted for  in the last step in a weighted DiD regression (equation \ref{eq:rcsdid}). Thus, those weights only need to make the outcomes from the treated and control groups parallel and not an exact match. Those weights $\homega^{sdid}$ are computed as:
\begin{equation}
\label{indiv_weights}
\begin{aligned}
& \left( \homega_0, \, \homega^\sdid \right) = \argmin_{ \omega_0 \in \mmr, \omega\in \Omega} \ell_{unit}(\omega_0, \omega) \quad \text{ where } \\
& \quad \quad \quad\ell_{unit}(\omega_0, \omega) = \sum_{t=1}^{T_\pre} 
\left( \omega_0 + \sum_{k=1}^{K_\ccc} \omega_{k} \bar{Y}_{k,t} -  
\frac{1}{K_\ttt} \sum_{k = K_\ccc + 1}^K \bar{Y}_{k,t}\right)^2 + \zeta^2 T_{\pre} \left\|\omega\right\|_2^2, \\
& \quad \quad \quad \Omega = \cb{\omega \in \mmr_+^N :   \sum_{k=1}^{K_\ccc} \omega_k=1, \,  \omega_k=K_\ttt^{-1} \text{ for all } k=K_\ccc+1, \, \ldots \, ,K},
\end{aligned}
\end{equation}
The regularization parameter $\zeta$ is:
\begin{equation}
\label{eq:penalization}
\begin{split}
&\zeta= (K_\ttt T_\post)^{1/4} \ \hat \sigma \ \text{ with }\  \hat \sigma^2 = \frac{1}{K_\ccc (T_\pre-1)}\sum_{k=1}^{K_\ccc}\sum_{t=1}^{T_\pre-1}  \left( \Delta_{k,t}-\overline{\Delta} \right) ^2, \\
&\mathrm{where}\ \  \Delta_{k,t}=\bar{Y}_{k,(t+1)}-\bar{Y}_{k,t},\hskip0.5cm {\rm and}\ \  \overline{\Delta}=\frac{1}{K_\ccc (T_\pre-1)}\sum_{k=1}^{K_\ccc}\sum_{t=1}^{T_\pre-1} \Delta_{k,t}.
\end{split}
\end{equation}
$\zeta$ allows increasing the dispersion of the weights and ensuring their uniqueness.\\

The time weights, $\hlambda^\sdid$, match pretreatment periods with post-treatment periods for the control group. They give more weight to pre-treatment periods that are similar to post-treatment periods. They are computed as:
\begin{equation}
\label{period_weights}
\begin{aligned}
& \left( \hlambda_0, \, \hlambda^\sdid \right) = \argmin_{ \lambda_0 \in \mmr, \lambda \in \Lambda}  \ell_{time}(\lambda_0, \lambda) \quad \text{ where } \\
&\quad \quad \quad \ell_{time}(\lambda_0, \lambda) = \sum_{k=1}^{K_\ccc} 
\left( \lambda_0 + \sum_{t=1}^{T_\pre} \lambda_{t} \bar{Y}_{k,t} -  
\frac{1}{T_\post} \sum_{t = T_\pre + 1}^T \bar{Y}_{k,t}\right)^2  , \\
&\quad \quad \quad \Lambda=\cb{\lambda\in\mmr_+^T :  \sum_{t=1}^{T_\pre} \lambda_t=1, \, \lambda_t=T_\post^{-1} \text{ for all } t=T_\pre+1,\, \ldots \, ,T}.
\end{aligned}
\end{equation}

Because the number of observations in each group may be different for each group-period, $\homega_k^\sdid$ do not ensure a similar trend between the treated and control groups. I compute cross-sectional weights that accounts for the different number of observations in each group-period. The cross-sectional weights $\nu_{k,t}^{RC}$ are computed as:
\begin{equation}
\label{eq:rc_weights}
\nu_{k,t}^{RC} = \frac{1}{N_{k,t}}
\end{equation}

As demonstrated in appendix \ref{sec:demowei}, using the cross-sectional weights $\nu_{k,t}^{RC}$ allow the total weights of each group-period to be equal to $\homega_k^\sdid \times \hlambda_t^\sdid$ as in the SDiD method.\\

Once the weights $\nu_{k,t}^{RC}$ are obtained, the treatment effect is estimated in a weighted regression as:
\begin{equation}
\label{eq:rcsdid}
\left( \hat\tau^{RC-SDiD}, \, \hat\mu, \, \hat\alpha, \, \hat\beta \right)=
\argmin_{\tau,\mu,\alpha,\beta}  \cb{ \sumit \left( Y_{i(k,t), t}-\mu-\alpha_k-\beta_t- \tau W_{k,t} \right) ^2\homega_k^\sdid\hlambda_t^\sdid \nu_{k,t}^{RC}}
\end{equation}
where $\mu$ is an intercept, $\alpha_k$ a group fixed effect and $\beta_t$ a time fixed effect. $W_{k,t}$ is a binary variable representing treatment exposure, and $\tau$ is the treatment effect.\\

The RC-SDiD method can be adapted to include covariates (\citealp{arkhangelsky_synthetic_2021, kranz_synthetic_2022}) and to staggered treatment timing (\citealp{arkhangelsky_synthetic_2021, porreca_synthetic_2022}).

\section{Monte Carlo simulations}
\label{sec:simul}
I use Monte Carlo simulations to study the properties of the RC-SDiD estimator compared to the SDiD and DiD estimators\footnote{The code can be found here: \url{https://github.com/yoannmorin/RC-SDID}}. I generate the outcome using a data generating process (DGP) similar to \citet{xu_generalized_2017} but without covariates. The unobservable dimension is generated using interactive fixed effects (\citealp{bai_panel_2009}) at the group level. The outcome for observation $i$ in group $k$ at time $t$ is simulated as:
\begin{equation}
\label{eq:ife}
Y_{i(k,t), t} =  \tau W_{k,t} + \alpha_k + \beta_t + \Lambda_k^{'} f_t + \varepsilon_{i(k,t), t}
\end{equation}
Only the first group is considered to be treated to keep a reasonable number of observations. The error term $\varepsilon_{i(k,t), t}$, time fixed effects $\beta_t$, and $f_t$ are i.i.d. $N(0,1)$. The values of group fixed effects $\alpha_k$ and factor loadings $\Lambda_k$ are drawn from uniform distributions. For the control groups, I use $U[-\sqrt{3}, \sqrt{3}]$ and $U[\sqrt{3} - 2w\sqrt{3}, 3\sqrt{3} - 2w\sqrt{3}] $ for the treated group, where $w \in [0,1]$. As highlighted in \citet{xu_generalized_2017}, it allows treatment status and group-specific effects to be correlated and to have a variance of 1 for the random variables (when $0\leq w <1$).\\

The varying number of observations in each group-period is computed as:
\begin{equation}
\label{eq:rceq}
N_{k,t} = \left\{
    \begin{array}{ll}
        S_k \times Base_{RC}  + S_k \times E_{k,t}    & \text{ if t=1} \\
        N_{k,t-1} + S_k \times E_{k,t}		& \text{ if t>1}
    \end{array}
\right.
\end{equation} 
where $S_k$ is a scale parameter for each group $k$. It allows the number of observations to differ between each group. It is drawn from a discrete uniform distribution and I test different range for this parameter. The draw of $S_k$ is allowed to be correlated with group fixed effects for different correlation levels $\rho$. $Base_{RC}$ is the baseline number of observations in each cross-section. To simulate the evolution of the number of observations, the variable $E_{k,t}$ is defined as:
\begin{equation}
\label{eq:rcevo}
E_{k,t} = N(Base_{RC} \times 0.02, \frac{\sqrt{Base_{RC}}}{2})
\end{equation} 
On average, the number of observations in each cross-section will slightly increase, while keeping a reasonable number of observations.\\

I draw new independent values of $\varepsilon_{i(k,t), t}$ for each of the 1000 repetitions. The other parameters are fixed, with $Base_{RC}=100$, $S_k \in [1,10]$ is correlated with group fixed effects for $\rho=0.2$, and $w=0.2$. I use 30 control groups and 30 periods where half are pre-treatment periods. The treatment effect is set to $\tau=0.3$ and the number of factors and loadinds is $r=1$. I compare the performance of the RC-SDiD estimator with DiD and SDiD by computing the mean bias, the standard deviation and the RMSE.\\

\begin{table}[htbp!]
\centering 
\begin{footnotesize}
\begin{tabular}{l S[table-format=2.4] S[table-format=2.4] S[table-format=2.4] S[table-format=2.4] S[table-format=2.4] S[table-format=2.4] S[table-format=2.4] S[table-format=2.4] S[table-format=2.4]}
\toprule
	&  \multicolumn{3}{c}{Mean bias} & \multicolumn{3}{c}{SD} & \multicolumn{3}{c}{RMSE} \\
\cmidrule{2-10}
	& \multicolumn{1}{c}{DiD} & \multicolumn{1}{c}{RC-SDiD} & \multicolumn{1}{c}{SDiD} & \multicolumn{1}{c}{DiD} & \multicolumn{1}{c}{RC-SDiD} & \multicolumn{1}{c}{SDiD} & \multicolumn{1}{c}{DiD} & \multicolumn{1}{c}{RC-SDiD} & \multicolumn{1}{c}{SDiD} \\
\midrule
$S_k=1$ & 0.4468401 & 0.0012772 & -0.0372557 & 0.0319505 & 0.0411381 & 0.0436336 & 0.4479798 & 0.0411374 & 0.0573583\\
$S_k \in [1,2]$ & 0.4472078 & 0.0004750 & -0.0437201 & 0.0231139 & 0.0308517 & 0.0331879 & 0.4478041 & 0.0308399 & 0.0548797\\
$S_k \in [1,4]$ & 0.4408936 & 0.0006512 & -0.0421817 & 0.0162026 & 0.0213472 & 0.0237602 & 0.4411909 & 0.0213465 & 0.0484074\\
$S_k \in [1,6]$ & 0.4349744 & -0.0013078 & -0.0445587 & 0.0128466 & 0.0172892 & 0.0200685 & 0.4351639 & 0.0173299 & 0.0488653\\
$S_k \in [1,8]$ & 0.4397171 & -0.0008328 & -0.0442612 & 0.0115770 & 0.0157521 & 0.0186335 & 0.4398693 & 0.0157662 & 0.0480199\\
$S_k \in [1,10]$ & 0.4379900 & 0.0010044 & -0.0433893 & 0.0102691 & 0.0129780 & 0.0165539 & 0.4381102 & 0.0130103 & 0.0464369\\
$S_k \in [1,15]$ & 0.4366106 & 0.0004564 & -0.0437629 & 0.0086050 & 0.0112326 & 0.0145466 & 0.4366953 & 0.0112363 & 0.0461149\\
$S_k \in [1,20]$ & 0.4374794 & 0.0001533 & -0.0444513 & 0.0079285 & 0.0099395 & 0.0128890 & 0.4375511 & 0.0099358 & 0.0462804\\
\bottomrule
\end{tabular}
\end{footnotesize}
\caption{Variation of the scale parameter}
\label{tab:scale}
\end{table} 

Table \ref{tab:scale} displays the effect of the scale parameter on the performance of the model. When the value of this parameter increases, the number of observations in the dataset also increases. Thus, an efficient estimate should perform better when $S_k$ grows. The DiD estimate is the most biased estimate because of interactive fixed effects, while the RC-SDiD is the least biased one. Its bias is reduced when $S_k$ grows. The SD of all models gets lower as $S_k$ grows because of the higher number of observations. However, the RMSE of the DiD and SDiD estimators are only slightly affected by the increase of $S_k$, while the RMSE of RC-SDiD is greatly improved. Thus, adding the third type of weight presented in section \ref{sec:rcsdid} significantly improves the performance of the estimator when the number of observations differs in each cross-section.\\

\begin{table}[htbp!]
\centering 
\begin{footnotesize}
\begin{tabular}{l S[table-format=2.4] S[table-format=2.4] S[table-format=2.4] S[table-format=2.4] S[table-format=2.4] S[table-format=2.4] S[table-format=2.4] S[table-format=2.4] S[table-format=2.4]}
\toprule
	&  \multicolumn{3}{c}{Mean bias} & \multicolumn{3}{c}{SD} & \multicolumn{3}{c}{RMSE} \\
\cmidrule{2-10}
	& \multicolumn{1}{c}{DiD} & \multicolumn{1}{c}{RC-SDiD} & \multicolumn{1}{c}{SDiD} & \multicolumn{1}{c}{DiD} & \multicolumn{1}{c}{RC-SDiD} & \multicolumn{1}{c}{SDiD} & \multicolumn{1}{c}{DiD} & \multicolumn{1}{c}{RC-SDiD} & \multicolumn{1}{c}{SDiD} \\
\midrule
$r=0$ & 0.0000435 & -0.0002061 & -0.0001183 & 0.0105941 & 0.0128497 & 0.0127793 & 0.0105889 & 0.0128449 & 0.0127734\\
$r=1$ & 0.4374556 & 0.0001486 & -0.0445688 & 0.0103890 & 0.0132753 & 0.0162685 & 0.4375788 & 0.0132695 & 0.0474424\\
$r=2$ & -0.0749635 & 0.0002476 & -0.0254401 & 0.0102829 & 0.0160906 & 0.0262904 & 0.0756647 & 0.0160845 & 0.0365745\\
$r=3$ & 0.9438644 & -0.0001313 & -0.0471673 & 0.0103632 & 0.0184334 & 0.0269487 & 0.9439212 & 0.0184246 & 0.0543163\\
$r=4$ & 1.6734638 & -0.0003758 & -0.0973059 & 0.0103222 & 0.0187815 & 0.0289441 & 1.6734957 & 0.0187758 & 0.1015154\\
\bottomrule
\end{tabular}
\end{footnotesize}
\caption{Variation of the number of factors and loadings}
\label{tab:ife}
\end{table} 

Table \ref{tab:ife} focuses on the effects of the number of factors and loadings on the performance of the estimators. When $r=0$, the DGP only includes group and time fixed effect, and the DiD estimator performs the best across all measures. However, when $r$ increases, it becomes the estimator that performs the worst because it cannot account for interactive fixed effects. The RC-SDiD estimator has the best performance across all values of $r>0$. The bias, SD and RMSE of the SDiD estimator are all highly increasing with $r$, while they stay moderate for the RC-SDiD one.\\

In appendix \ref{sec:simw}, \ref{sec:simrho} and \ref{sec:simn} the results are reported for respectively: different values of $w$, different values of $\rho$ and for different sample sizes. In all cases, RC-SDiD performs better than DiD and SDiD, confirming our previous results.

\section{Conclusion}\label{sec:conclusion}
In this paper, I adapted the SDiD estimator (\citealp{arkhangelsky_synthetic_2021}) for repeated cross-sectional data, with a simple implementation that consists in adding a third type of weight that depends on the number of observations in each pair of group-period. Using Monte Carlo simulations, I demonstrated that the RC-SDiD estimator significantly improves the performance of the SDiD estimator when using repeated cross-sectional data.

\section*{Acknowledgements}
I am grateful to the comments and suggestions from Marie Breuillé, Capucine Chapel, Julie Le Gallo and Morgan Ubeda. Financial support from the Atelier Parisien d'Urbanisme is gratefully acknowledged.

\newpage
\appendix
\renewcommand{\thesection}{\Alph{section}.\arabic{section}}
\setcounter{section}{0}
\setcounter{table}{0}
\setcounter{equation}{0}
\counterwithin{figure}{section}
\counterwithin{table}{section}
\counterwithin{equation}{section}
\renewcommand\thefigure{\thesection\arabic{figure}}
\renewcommand\thetable{\thesection\arabic{table}}
\renewcommand\theequation{\thesection\arabic{equation}}

\begin{appendices}

\section{Cross-sectional weights}\label{sec:demowei}
For a group $k$ at time $t$, the weight computed using the SDID method is:
\begin{equation}\label{eq:wkt}
W_{k,t} = \omega_k \times \lambda_t
\end{equation}

Because we have multiple observations in each group-period $(k,t)$, the total weight for a group-period is:
\begin{equation}\label{eq:sumwkt}
\sum_i W_{k,t} = \omega_k \times \lambda_t \times N_{k,t}
\end{equation}
Thus, estimating the weighted difference-in-differences regression would lead to biased results because the actual weights  are different than the ones computed by the SDID method.

The only case where the estimator would not be biased is if $N_{k,t} = N_{p,c}$, $\forall k \neq p \text{ and } t \neq c$, then the relative weights of each groups are equal.

However, if $N_{k,t} \neq N_{p,c}$, we need to apply the weights suggested in equation \ref{eq:rc_weights}. Using those weights, the total weight for each pair $(k,t)$ is:
\begin{equation}\label{eq:sumwkt_correct}
\sum_i W_{k,t} = \omega_k \times \lambda_t \times N_{k,t} \times \frac{1}{N_{k,t}} =  \omega_k \times \lambda_t
\end{equation}
The weights correspond to those in equation \ref{eq:wkt} and allow to have unbiased estimates of the treatment effect using the weighted difference-in-differences regression.

\newpage
\section{Additional simulation results}
\subsection{Treatment assignment}\label{sec:simw}
Table \ref{tab:w} presents the results for different values of $w$. If $w=1$, treatment assignment is random, while when it decreases toward 0, individual fixed effects and factor loadings are more shifted, reducing the overlap between the control and treated groups. As in the previous table, the RC-SDiD estimator performs the best in all cases. The more the treatment status is correlated with individual effects, the more the SDiD and DiD estimators are biased. The RC-SDiD estimator performs well for all values of $w$, and its SD and RMSE are only slightly higher for the highest values of $w$ compared to the lowest ones. \\

\begin{table}[htbp!]
\centering 
\begin{footnotesize}
\begin{tabular}{l S[table-format=2.4] S[table-format=2.4] S[table-format=2.4] S[table-format=2.4] S[table-format=2.4] S[table-format=2.4] S[table-format=2.4] S[table-format=2.4] S[table-format=2.4]}
\toprule
	&  \multicolumn{3}{c}{Mean bias} & \multicolumn{3}{c}{SD} & \multicolumn{3}{c}{RMSE} \\
\cmidrule{2-10}
	& \multicolumn{1}{c}{DiD} & \multicolumn{1}{c}{RC-SDiD} & \multicolumn{1}{c}{SDiD} & \multicolumn{1}{c}{DiD} & \multicolumn{1}{c}{RC-SDiD} & \multicolumn{1}{c}{SDiD} & \multicolumn{1}{c}{DiD} & \multicolumn{1}{c}{RC-SDiD} & \multicolumn{1}{c}{SDiD} \\
\midrule
$w=1$ & -0.1408411 & -0.0001588 & 0.0108149 & 0.0112322 & 0.0129366 & 0.0131745 & 0.1412878 & 0.0129311 & 0.0170398\\
$w=0.8$ & 0.0038626 & 0.0000360 & 0.0025244 & 0.0110672 & 0.0122658 & 0.0123542 & 0.0117167 & 0.0122597 & 0.0126034\\
$w=0.6$ & 0.1485963 & 0.0004696 & -0.0071239 & 0.0103486 & 0.0120534 & 0.0125895 & 0.1489559 & 0.0120566 & 0.0144599\\
$w=0.4$ & 0.2931235 & 0.0006213 & -0.0201137 & 0.0102610 & 0.0122485 & 0.0138068 & 0.2933029 & 0.0122582 & 0.0243926\\
$w=0.2$ & 0.4372114 & 0.0000452 & -0.0442913 & 0.0107057 & 0.0138816 & 0.0168912 & 0.4373423 & 0.0138747 & 0.0473999\\
$w=0$ & 0.5818943 & -0.0001030 & -0.0704880 & 0.0103791 & 0.0146666 & 0.0204045 & 0.5819868 & 0.0146597 & 0.0733791\\
\bottomrule
\end{tabular}
\end{footnotesize}
\caption{Variation of treatment assignment}
\label{tab:w}
\end{table}

\newpage
\subsection{Correlation between fixed effects and cross-section size}\label{sec:simrho}
The results for different values of $\rho$ are presented in table \ref{tab:cors}. This parameter controls the correlation between the scale parameter $S_k$ and the individual fixed effects. This parameter has overall a pretty low effect on the simulation results. The mean bias, standard deviation and RMSE all have similar values for the different values of $\rho$ that are tested. The RC-SDID estimates are on average the least biased and have the lowest RMSE. Its standard deviation is only slightly higher than the DID one (which is highly biased) and is always lower than for SDID estimates.

\begin{table}[htbp!]
\centering 
\begin{footnotesize}
\begin{tabular}{l S[table-format=2.4] S[table-format=2.4] S[table-format=2.4] S[table-format=2.4] S[table-format=2.4] S[table-format=2.4] S[table-format=2.4] S[table-format=2.4] S[table-format=2.4]}
\toprule
	&  \multicolumn{3}{c}{Mean bias} & \multicolumn{3}{c}{SD} & \multicolumn{3}{c}{RMSE} \\
\cmidrule{2-10}
	& \multicolumn{1}{c}{DiD} & \multicolumn{1}{c}{RC-SDiD} & \multicolumn{1}{c}{SDiD} & \multicolumn{1}{c}{DiD} & \multicolumn{1}{c}{RC-SDiD} & \multicolumn{1}{c}{SDiD} & \multicolumn{1}{c}{DiD} & \multicolumn{1}{c}{RC-SDiD} & \multicolumn{1}{c}{SDiD} \\
\midrule
$\rho=0$ & 0.4253999 & -0.0005093 & -0.0378913 & 0.0119574 & 0.0155577 & 0.0196384 & 0.4255678 & 0.0155583 & 0.0426735\\
$\rho=0.2$ & 0.4373224 & -0.0000620 & -0.0446836 & 0.0099866 & 0.0138605 & 0.0166467 & 0.4374363 & 0.0138537 & 0.0476808\\
$\rho=0.5$ & 0.4465070 & -0.0002270 & -0.0421138 & 0.0158551 & 0.0196222 & 0.0247974 & 0.4467881 & 0.0196137 & 0.0488658\\
$\rho=0.8$ & 0.4579425 & -0.0002093 & -0.0423577 & 0.0125893 & 0.0163280 & 0.0203522 & 0.4581154 & 0.0163211 & 0.0469891\\
$\rho=1$ & 0.4522672 & -0.0008566 & -0.0419693 & 0.0132856 & 0.0170992 & 0.0209437 & 0.4524621 & 0.0171121 & 0.0469001\\
\bottomrule
\end{tabular}
\end{footnotesize}
\caption{Variation of the correlation between individual fixed effects and the scale parameter}
\label{tab:cors}
\end{table}

\newpage
\subsection{Sample size}\label{sec:simn}
The results reported in table \ref{tab:size} study the effects of sample size on the performance of the three estimators. Samples are simulated with a different number of baseline observations in each cross-section (50 vs 100) and a different number of control groups and periods (30 vs 15 for both). For each sample size considered, RC-SDiD still performs better than DiD and SDiD estimates. The performance of the RC-SDiD estimator is more impacted by reducing the number of time periods than the number of groups. Reducing the number of observations in each cross-section also slightly raise its RMSE and standard deviation. But it performs better than DiD and SDiD in all cases.

\begin{table}[htbp!]
\centering 
\begin{footnotesize}
\begin{tabular}{l S[table-format=2.4] @{\hspace{0.4\tabcolsep}} S[table-format=2.4] @{\hspace{0.4\tabcolsep}} S[table-format=2.4] @{\hspace{0.4\tabcolsep}} S[table-format=2.4] @{\hspace{0.4\tabcolsep}} S[table-format=2.4] @{\hspace{0.4\tabcolsep}} S[table-format=2.4] @{\hspace{0.4\tabcolsep}} S[table-format=2.4] @{\hspace{0.4\tabcolsep}} S[table-format=2.4] @{\hspace{0.4\tabcolsep}} S[table-format=2.4] @{\hspace{0.4\tabcolsep}}}
\toprule
	&  \multicolumn{3}{c}{Mean bias} & \multicolumn{3}{c}{SD} & \multicolumn{3}{c}{RMSE} \\
\cmidrule{2-10}
	& \multicolumn{1}{c}{DiD} & \multicolumn{1}{c}{RC-SDiD} & \multicolumn{1}{c}{SDiD} & \multicolumn{1}{c}{DiD} & \multicolumn{1}{c}{RC-SDiD} & \multicolumn{1}{c}{SDiD} & \multicolumn{1}{c}{DiD} & \multicolumn{1}{c}{RC-SDiD} & \multicolumn{1}{c}{SDiD} \\
\midrule
$Base_{RC}=100$\\
\cmidrule{1-1}
$K^\ccc=30, T=30$  & 0.4368285 & -0.0002501 & -0.0443700 & 0.0105341 & 0.0135357 & 0.0165137 & 0.4369554 & 0.0135312 & 0.0473405\\
$K^\ccc=15, T=30$ & 0.4542206 & -0.0003190 & -0.0691387 & 0.0102860 & 0.0149355 & 0.0225311 & 0.4543369 & 0.0149315 & 0.0727139\\
$K^\ccc=30, T=15$ & -0.1762892 & -0.0001341 & -0.0408285 & 0.0153515 & 0.0186173 & 0.0196246 & 0.1769557 & 0.0186085 & 0.0452957\\
$K^\ccc=15, T=15$ & -0.1818026 & 0.0006477 & -0.0455559 & 0.0157804 & 0.0206981 & 0.0225544 & 0.1824855 & 0.0206979 & 0.0508285\\
\midrule
$Base_{RC}=50$\\
\cmidrule{1-1}
$K^\ccc=30, T=30$ & 0.4304450 & -0.0010989 & -0.0468628 & 0.0143430 & 0.0193453 & 0.0217074 & 0.4306837 & 0.0193668 & 0.0516417\\
$K^\ccc=15, T=30$ & 0.4449829 & 0.0008558 & -0.0745307 & 0.0143064 & 0.0219210 & 0.0305070 & 0.4452126 & 0.0219267 & 0.0805269\\
$K^\ccc=30, T=15$ & -0.1668488 & -0.0003760 & -0.0415113 & 0.0233776 & 0.0283635 & 0.0288067 & 0.1684769 & 0.0283518 & 0.0505191\\
$K^\ccc=15, T=15$ & -0.1728540 & -0.0017580 & -0.0479654 & 0.0224712 & 0.0317394 & 0.0338980 & 0.1743070 & 0.0317722 & 0.0587248\\
\bottomrule
\end{tabular}
\end{footnotesize}
\caption{Variation of the number of control groups, time periods, and baseline number of observation by group}
\label{tab:size}
\end{table}

\end{appendices}

\end{document}